\documentclass[nofootinbib,preprint]{revtex4}
\usepackage{epsfig}
\newcommand{\be}{\begin{eqnarray}}\newcommand{\ee}{\end{eqnarray}}
\newcommand{\beo}{\begin{eqnarray*}}\newcommand{\eeo}{\end{eqnarray*}}
\newcommand{\no}{\nonumber}
\begin{document}
\preprint{APCTP 2001-004}
\baselineskip=18pt
\parskip=3pt

\title{Boundary String Field Theory at One-loop}

\author{Taejin Lee}
\email{taejin@cc.kangwon.ac.kr} \affiliation{Department of
Physics, Kangwon National University, Chuncheon 200-701, Korea \\
Asia Pacific Center for Theoretical Physics POSTECH, 790-784,
Korea}
\author{K. S. Viswanathan}
\email{kviswana@sfu.ca}
\author{Yi Yang}
\email{yyangc@sfu.ca} \affiliation{Department of Physics, Simon
Fraser University, Burnaby, BC V5A 1S6 Canada}
\date{\today}

\begin{abstract}
We discuss the open string one-loop partition function in tachyon
condensation background of a unstable D-brane system. We evaluate
the partition function by using the boundary state formulation and
find that it is in complete agreement with the result obtained in
the boundary string field theory. It suggests that the open string
higher loop diagrams may be produced consistently by a closed
string field theory, where the D-brane plays a role of source for
the closed string field.
\end{abstract}

%04.60.Ds Canonical quantization
%11.25.-w Theory of fundamental strings
%11.25.Sq Nonperturbative techniques; string field theory
%11.25.H Algebraic methods in string theory
%11.27 Strings extended classical solutions, 

\maketitle
\thispagestyle{empty}

\section{Introduction}
The boundary string field theory (BSFT) \cite{witten92,shata93}
has been proved to be a useful tool to study the dynamics of the
D-branes in string theory. It is this framework where the tachyon
condensation on unstable D-brane systems has been discussed
extensively in recent works \cite{harvey00,noncomm,andreev}. The
BSFT deals with the disk partition function of the open string
theory. The relationship between the BSFT action, $S$ and the disk
partition function $Z$ is given as \be S = \left( 1+ \beta^i
\frac{\partial}{\partial g^i} \right) Z \ee where $g^i$ are the
couplings of the boundary interactions and $\beta^i$ are the
corresponding world-sheet $\beta$-functions. If a specific form of
tachyon profile is chosen, the BSFT action reduces to the
effective action for the tachyon field. The BSFT has been extended
to the case of superstring theories \cite{super}, where the BSFT
action coincides with the disk partition function $Z$. The results
obtained in those studies confirm Sen's conjecture; as the tachyon
condensation develops, the unstable open string vacuum
spontaneously rolls down to the stable closed string vacuum.

The open string one-loop partition function is also studied
recently in refs. \cite{vish1,vish2,one-loop,craps}. Assuming that
the BSFT action may coincide with the partition function beyond
the disk diagram in superstring theory, we find that the effective
action for tachyon may receive corrections from the open string
loop diagrams. The effective action for tachyon at the lowest
order is obtained from the disk partition function with a specific
form of the tachyon profile on the boundary of the disk. In the
string theory we expect that there are some corrections to the
disk partition function. While keeping fixed the boundary of the
disk and the tachyon profile on it, we may add more boundaries and
handles to the string world-sheet diagram. They may give
corrections to the disk partition function, thus the effective
action for the tachyon field. We may propose that the total
partition function may be written schematically as \be
\label{total} Z = \sum_{b=1,\, g=0} \int d m Z[b,g;m] \ee where
$b$ denotes the number of boundaries and $g$, the number of
handles. We collectively denote the moduli parameters by $[m]$.
One of the boundaries of the string world sheet, on which the
partition function is evaluated, is fixed as in the disk diagram.
%Accordingly the effective potential for the
%tachyon may receive corrections from higher-loop diagrams.

\begin{figure}
\epsfxsize=11cm \centerline{\epsffile{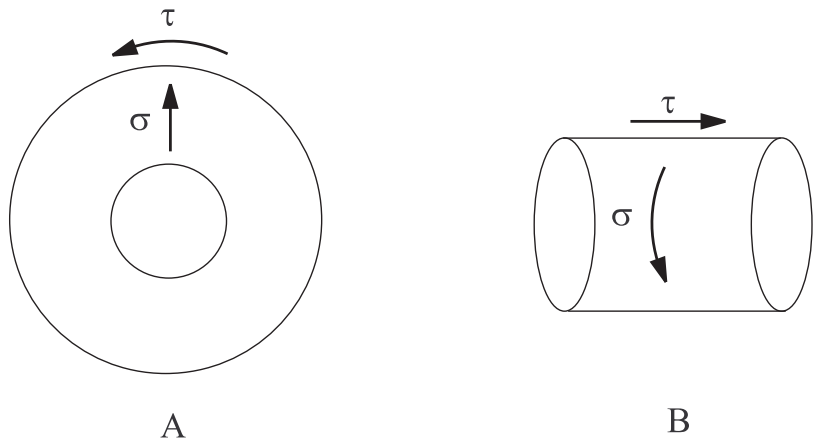}} \vspace{0cm}
\caption{A. Open string channel. B. Closed string channel.}
\label{open-closed}
\end{figure}

The leading correction to the disk partition function may come
from the open string one-loop diagram which has two boundaries.
One may employ two different schemes to calculate the one-loop partition
function. One is to use the open string channel, which adopts the
open string Green function on an annulus \cite{vish1,vish2,one-loop},
and the other is to use the closed string channel
\cite{tlee1,itoyama,semen}, which adopts the
boundary state formulation \cite{callan}. In the second scheme the
geometry of the string world sheet is given by a cylinder. A of
figure \ref{open-closed} depicts the one-loop diagram in the open
string channel. If one once calculates the Green function on the
annulus, the partition function is readily obtained as
\be\label{partition}
\frac{d}{dy}\ln
Z(u,a)=-\frac{1}{8\pi}\int_0^{2\pi}d\sigma\langle X^2\rangle,
\quad
Z = \int \frac{da}{a} Z(u,a)
\ee
where $u$ is the tachyon profile parameter and $a$ is the modular
on a cylinder. As
depicted by B of figure \ref{open-closed}, the same diagram can
be viewed in the closed string channel as follows; a closed string
appears from a D-brane and propagates to another (or the same)
D-brane. The initial and the final states of the closed string are
described by boundary states $|B \rangle$ and the one-loop
partition function is obtained as
\be\label{cor}
Z(u,\tau)=\langle
B|\exp\left[-\tau(L_0+\tilde{L}_0)\right]|B\rangle, \quad
Z = \int d\tau Z(u,\tau) .
\ee
As is well known, the partition functions obtained in
two different schemes agree on-shell thanks to the open-closed string
duality. However, it remains to be checked whether two different schemes
yield the same result in the presence of the tachyon condensation,
i.e., off-shell. Since the boundary state formulation entirely depends
on this equivalence, it is important to clarify this issue.
At the tree level, it has been shown that the disk partition
function in the BSFT appears as the normalization factor of the
boundary state. As we shall see that the open string one-loop
partition function calculated in the boundary state formulation,
Eq.(\ref{cor}) is in complete agreement with that obtained in the
BSFT, Eq.(\ref{partition}). It suggests that the open string
partition functions may be reproduced in a closed
string field theory, where the D-brane plays a role of source for
the closed string field. Constructing such a theory, we may have
a consistent framework to calculate the total partition function
Eq.(\ref{total}) in a systematic way, thus the corrections to
the effective action for tachyon.

In the next section, we discuss the bosonic string theory at open
string one-loop in the background of the tachyon condensation and show
by an explicit calculation that two different schemes yield the same
result. In section 3, we extend it to the supersymmetric theory.
Section 4 concludes the paper with some discussions.

\section{Bosonic String Field Theory at One-loop}

We start with the action
\be\label{Lagrangian}
S &=& S_{\rm bulk}+S_{\rm bndy} \no\\
&=& \frac{1}{8\pi}\int_\Sigma
d^2x\sqrt{h}h^{ij}\partial_iX^\mu\partial_jX_\mu +
\frac{1}{8\pi}\int_{\partial\Sigma}d\sigma \mathcal{V}.
\ee
$S_{\rm bndy}$ describes the interactions between the string and
the background, which may be expanded as
\be
\mathcal{V}=T(X)+A_\mu(X)\partial_\sigma
X^\mu+C_{\mu\nu}\partial_\sigma X^\mu\partial_\sigma X^\nu+\cdots.
\ee
In this paper, we only consider the tachyon field background,
which take the following simple form
\be
\mathcal{V}=T(X)\equiv\sum_{i=1}^pu_i(X^i)^2.
\ee

\noindent \bf{Open string channel - Annulus}\rm

At the tree level, the world sheet is a unit disk on which the
boundary action is given
\be\label{bndy}
S_{\rm bndy}=\frac{1}{8\pi}\int_0^{2\pi}d\sigma\,(uX^2).
\ee
Throughout this paper we are only concerned with the string
variables in the direction where the tachyon condensation
takes place. So we omit the space-time indices hereafter.
The geometry of the world sheet for the open string one-loop diagram
is an annulus with $a\leq r \leq b$, which has two boundaries.
It leads us to introduce the following boundary action
\be\label{boundary1}
S_{\rm bndy}=\frac{1}{8\pi}\left[\int_0^{2\pi}d\sigma\,(u_aX^2)|_a+
  \int_0^{2\pi}d\sigma\,(u_bX^2)|_b\right].
\ee One may consider an alternative form of the boundary action,
which makes use of the induced metric on the world sheet
\be\label{boundary2} S^\prime_{\rm
bndy}=\frac{1}{8\pi}\left[\int_0^{2\pi}a\, d\sigma\,(u_aX^2)|_a+
\int_0^{2\pi}b\,d\sigma\,(u_bX^2)|_b\right]. \ee However, if we
take $S^\prime_{\rm bndy}$ as the boundary action, we would
encounter difficulties such as the breakdown of the
Fischler-Susskind mechanism \cite{fischler} as pointed out in ref.
\cite{craps}. Hence, the theory may become singular. If we turn on
the tachyon condensation, the system is no longer conformally
invariant. But we expect that the system is invariant under the
following transformation, $z\rightarrow ab/\bar z$, which maps the
inner boundary of the annulus onto the outer boundary and vice
versa. It is obvious that $S^\prime_{\rm bndy}$ does not respect
this symmetry while $S_{\rm bndy}$ does. For these reasons we take
$S_{\rm bndy}$ Eq.(\ref{boundary1}) as the boundary action. As we
shall see, the partition function with the boundary action $S_{\rm
bndy}$ in the open string channel agrees with that obtained in the
boundary state formulation.

It follows from the action (\ref{Lagrangian}) that the world sheet
Green's function
\be
\triangle_zG(z,w)=-4\pi\delta^{(2)}(z-w),\ee
satisfies the following conditions
\be\label{bc}
\left(a\frac{\partial}{\partial
r}-u_a\right)G_B(z,w)|_{r=a}&=&0,\no\\
\left(b\frac{\partial}{\partial
r}+u_b\right)G_B(z,w)|_{r=b}&=&0, \ee
which read in the complex coordinates
$(z,\bar z)=(re^{i\sigma},re^{-i\sigma})$ as
\be
\partial_z\bar\partial_{\bar z}G(z,w)=-2\pi\delta^{(2)}(z-w),
\ee
with the boundary conditions
\be (z\partial +\bar z\bar\partial
-u_a)G_B(z,w)|_{r=a}&=&0,\no\\ (z\partial +\bar z\bar\partial
+u_b)G_B(z,w)|_{r=b}&=&0.
\ee

An explicit form of the Green function is obtained by
solving this boundary problem \cite{vish1}. The partition
function is then determined by Eq.(\ref{partition}) up to a integration
constant $c_B$
\be
Z_B(u_a,u_b;a,b)&=& c_B T^2_p \left[u_a+u_b-\frac{u_au_b}{2} \ln
\left(\frac{a^2}{b^2}\right)\right]^{-1/2}\no\\ &
& \prod_{n=1}^\infty\left[(n+u_a)(n+u_b)
-(n-u_a)(n-u_b)\left(\frac{a}{b}\right)^{2n}\right]^{-1}
\ee
where $T_p$ is the tension of the Dp-brane.
Here we set $\alpha^\prime = 2$.
We consider the case where both ends of the open string are
attached on the same D-brane, $u_a=u_b=u$. Since we are concerned
with the corrections to the disk partition function, one of the
boundaries should be same as the boundary of the disk. Thus,
choosing $b=1$, we obtain
\be\label{open}
Z_B(u,a)= c_B T^2_p \sqrt{\frac{1}{2u-u^2\ln
a}}\prod_{n=1}^\infty\frac{1}{(n+u)^2 -(n-u)^2a^{2n}}.
\ee

\noindent {\bf Closed string channel - Cylinder}

In the boundary state formulation the open string one-loop diagram
is depicted as a cylinder diagram, where
a closed string comes out from the D-brane and propagates onto the
D-brane. The interaction between the D-brane background and the
string is encoded in the initial and final states of the closed
string, which are termed the boundary states. Construction of the
boundary state begins with defining the boundary states
$\{|X \rangle \}$ which form a basis for the closed string states \be
\hat{X} |X \rangle &=& X |X \rangle, \\
\hat{X}(\sigma) &=& \hat{x}_0 + \sum_{n=1} \frac{1}{\sqrt{n}}
\left\{(a_n + \tilde{a}^{\dagger}_n) e^{2in\sigma} +
(a^{\dagger}_n +\tilde{a}_n) e^{-2in\sigma} \right\} \no\\
X(\sigma) &=& x_0 + \sum_{n=1} \frac{1}{\sqrt{n}} \left(x_n
e^{2in\sigma} + \bar{x}_n e^{-2in\sigma}\right) \no \ee where
$\sigma \in [0, \pi]$ and $[a_n, a^\dagger_m] = [\tilde{a}_n,
\tilde{a}^\dagger_m] = \delta_{nm}$. For a given boundary action
$S_{\rm bndy}$ the boundary state is constructed as \be |B \rangle
= T_p \int D[x, \bar{x}] e^{i S_{\rm bndy}[x,\bar{x}]} |x, \bar{x}
\rangle. \ee
Choosing $S_{\rm bndy}$ Eq.(\ref{bndy}) as the
boundary action in order to describe the tachyon background we
obtain the boundary state \cite{tlee1} as \be\label{bb}
|B_B\rangle&=& T_p \prod_{n=1}\left(1+\frac{u}{n}\right)^{-1}\int
dxe^{-\frac{1}{4} x u x}\exp{\left\{a_n^\dag
\left(\frac{n-u}{n+u}\right)\tilde{a}_n^\dag\right\}|x\rangle} \ee
where $| x \rangle$ is defined as \be \hat{x}_0 |x \rangle = x |x
\rangle, \quad a^\dagger_n |x \rangle = \tilde{a}^\dagger_n |x
\rangle = 0, \ee where $n = 1, 2, 3, \dots$. (Note that the
tachyon profile parameter $u$ in the ref. \cite{tlee1} differs
from that in the ref. \cite{vish1,vish2} by $1/4\pi$.)

We can calculate the partition function by making use of
Eq.(\ref{cor})
\be\label{close}
Z_B(u,\tau)&=& T^2_p \langle B_B| \exp \left[-\tau(L_0 + \tilde{L}_0)
\right] |B_B\rangle \no\\
&=& T^2_p \prod_{n=1}\left(1+\frac{u}{n}\right)^{-2}\int dxdx'\no
\langle x'|e^{-\frac{1}{4} x'u x'}\exp{\left\{\tilde{a}_n
\left(\frac{n-u}{n+u}\right)a_n\right\}} \no\\
& & e^{-\tau(\hat{p}^2+ N + \tilde{N})}\exp{\left\{a_n^\dag
\left(\frac{n-u}{n+u}\right)\tilde{a}_n^\dag\right\}}e^{-\frac{1}{4}
x u x}|x\rangle\no\\
&=& T^2_p \sqrt{\frac{2\pi}{u+u^2\tau/2}}
\prod_{n=1}\left(1+\frac{u}{n}\right)^{-2}
\frac{1}{1-e^{-2n\tau}\left(\frac{n-u}{n+u}\right)^2}
\ee
where
\be
N = \sum_{n=1} n a^\dagger_n a_n, \quad \tilde{N} = \sum_{n=1}
n \tilde{a}^\dagger_n \tilde{a}_n. \no
\ee
Comparing the partition function Eq.(\ref{close}) in the closed string
channel with that Eq.(\ref{open}) in the open string channel and using
the zeta function regularization
\be
\prod_{n=1} \frac{1}{n +\epsilon} = \exp \left\{
\frac{d}{ds} \left(\zeta(s, \epsilon) - \epsilon^{-s}\right) \right\}_{s=0}
= \frac{\epsilon \Gamma(\epsilon)}{\sqrt{2\pi}},
\ee
we find that they coincide if we choose
\be
\tau=-\ln a, \quad c_B = 4\pi^{3/2}.
\ee

\section{Superstring Field Theory at One-loop}
In this section, we extend our previous discussion on equivalence
of two schemes to the superstring theory.
The bulk action in the supersymmetric theory is given as the
superstring world sheet action
\be
S_{\rm bulk}=\frac{1}{8\pi}\int_\Sigma d^2z\left(\partial
X^\mu\bar{\partial}X +\psi \bar{\partial}\psi +
\tilde{\psi}\partial\tilde{\psi}\right),
\ee
where $\psi$ and $\tilde{\psi}$ are the holomophic and
antiholomophic fermionic fields.
The interaction between the tachyon background and the superstring
is described by the following boundary action
\be
S_{\rm bndy}=\frac{1}{8\pi }\int_{\partial\Sigma}d\sigma
\left[(T(X))^2+(\psi\partial T) \frac{1}{\partial_\sigma}
(\psi\partial T)+(\tilde{\psi}\partial T)
\frac{1}{\partial_\sigma}(\tilde{\psi}\partial T)\right].
\ee
In order to produce the same tachyon profile in the bosonic
sector as in the bosonic string theory we choose
$T(X)=uX$. The total world sheet action for the superstring is
\be \label{superl}
S = S_{\rm bulk} + S_{\rm bndy}.
\ee

\noindent \bf{Open string channel - Annulus}\rm

The bosonic part is exactly the same as in the bosonic string
theory discussed in the last section. The superstring has two
different fermionic sectors, depending on the boundary conditions
for the fermion fields; R-R sector and NS-NS sector. In this
paper we only consider the NS-NS sector. The equations for the
fermion Green functions, following from the superstring
action Eq.(\ref{superl}) are
\be
\bar{\partial}G_F(z,w)&=&-i\sqrt{zw}\delta^{(2)} (z-w),\no\\
\partial\tilde{G}_F(\bar{z},\bar{w})&=&+i\sqrt{\bar{z}\bar{w}}
\delta^{(2)} (\bar{z}-\bar{w})
\ee
\be
\left.\left(1-iy_a\frac{1}{\partial_\sigma}\right)G_F\right|_{r=a}
&=&\left.\left(1+iy_a\frac{1}{\partial_\sigma}\right)
\tilde{G}_F\right|_{r=a},\no\\
\left.\left(1+iy_b\frac{1}{\partial_\sigma}\right)G_F\right|_{r=b}
&=&\left.\left(1-iy_b\frac{1}{\partial_\sigma}\right)\tilde{G}_F
\right|_{r=b}
\ee where $G(z,w)$ and $\tilde G(\tilde z,\tilde w)$ are the
holomophic and antiholomophic fermion Green function,
and $y\equiv u^2$.

Explicit expressions of the fermion Green functions can be found
in \cite{vish2}. Then a simple integration over $y$ yields the
partition function in the superstring theory Eq.(\ref{partition}).
Up to a constant $c_S$ the partition function is obtained as
\cite{vish2}\footnote{This expression differs by a factor $4^y$
from that of ref. \cite{vish2} due to a different choice of
regularization scheme. Similar to the disk case, this difference
does not affect any physical quantities. However, we may easily
remove this factor by choosing an alternative regularization
scheme as follows: In ref. \cite{vish2}, the divergences of
bosonic Green function and fermionic Green function cancel each
other and leave a constant $(-8\ln 2)$, which leads to the $4^y$
factor in the partition function. If we follow the regularization
scheme in \cite{witten92} \be \langle X(\theta)X(\theta)\rangle
|_{\partial\Sigma}= \lim_{\epsilon\to
0}\left[X(\theta)X(\theta)-\ln(1-e^{i\epsilon})-\ln(1-e^{-i\epsilon})
\right] \no \ee for both bosonic and fermionic Green functions
separately, we will obtain the expression (\ref{superopen}).}
\be\label{superopen} Z_S (y,a)&=& c_S
(2T_p)^2\sqrt{\frac{1}{2y-y^2\ln
{a}}}\prod_{n=1}^\infty\frac{(n+2y)^2 -(n-2y)^2a^n}
{\left[(n+y)^2-(n-y)^2a^{2n}\right]^2}. \ee The normalization of
the one-loop partition function taken here is consistent with
refs. \cite{super,ghoshal}.

\noindent \bf{Closed string channel - Cylinder}\rm

In the superstring theory the boundary state may be given as
$|B\rangle=|B_B \rangle\otimes|B_F\rangle$. Here $|B_B \rangle$,
describing the interaction between the bosonic degrees of freedom
of the superstring and the tachyon condensation background,
is already given  by Eq.(\ref{bb}). The interaction between the
fermion degrees of freedom and the tachyon condensation background
is encoded in $|B_F \rangle$. Construction of the fermionic part of
the boundary state is similar to that of the bosonic part.
It begins with defining the following eigenstate in the NS-NS sector
\be
\frac{1}{\sqrt{2}} \left(\psi + \tilde{\psi} \right)
|\theta, \bar{\theta} \rangle
= \sum_{r = 1/2} \left(\theta_r e^{2ir\sigma} + \bar{\theta}_r
e^{-2ir\sigma} \right) |\theta, {\bar \theta} \rangle .
\ee
Here
\be
\psi= \sqrt{2} \sum_{r=1/2} \left(b_r e^{2ir\sigma} +
b^\dagger_r e^{-2ir\sigma} \right), \no\\
\tilde{\psi}= \sqrt{2} \sum_{r=1/2} \left(\tilde{b}_r e^{2ir\sigma} +
\tilde{b}^\dagger_r e^{-2ir\sigma} \right), \no\\
\{b_r, b^\dagger_s \} = \delta_{rs}, \quad
\{\tilde{b}_r, \tilde{b}^\dagger_s \} = \delta_{rs}. \no
\ee
The fermionic part of the boundary state $|B_F \rangle$ is
written in terms of $|\theta, \bar{\theta} \rangle$ as
\be
|B_F \rangle = \int D[\theta, \bar{\theta}]
e^{S^F_{\rm bndy}[\theta, \bar{\theta}]} |\theta, \bar{\theta} \rangle .
\ee
For the background of the tachyon condensation we have
\be
S^F_{\rm bndy} &=& \frac{y}{4\pi} \int d\sigma \,
\theta (\partial_\sigma)^{-1} \theta \\
\theta (\sigma) &=& \sum_{r = 1/2} \left(\theta_r e^{2ir\sigma} +
\bar{\theta}_r e^{-2ir\sigma} \right). \no
\ee
A simple algebra yields the explicit form of $|B_F \rangle$
\be
|B_F \rangle=\prod_{r=\frac{1}{2}}\left(1+\frac{y}{r}\right)
\exp{\left\{b_r^\dag \left(\frac{r-y}{r+y}\right)\tilde{b}_r^\dag
\right\}|0_F\rangle}
\ee
where
\be
b^\dagger_r |0_F \rangle = \tilde{b}^\dagger_r
|0_F \rangle= 0. \no
\ee

As in the bosonic sector the contribution of the fermionic sector
to the partition function is written as the expectation value
of the superstring propagator
\be\label{closef}
Z_F(y,\tau)&=& \langle B_F|\exp \left[-\tau
(L^F_0 + \tilde{L}^F_0)\right]|B_F \rangle \\
&=&\prod_{r=\frac{1}{2}}\left(1+\frac{y}{r}\right)^2\langle 0_F|
\exp{\left\{\tilde{b}_r \left(\frac{r-y}{r+y}\right)b_r\right\}}
e^{-\tau(N_F + \tilde{N}_F)} \no \\
& & \quad \exp{\left\{b_r^\dag
\left(\frac{r-y}{r+y}\right)\tilde{b}_r^\dag\right\}}|0_F \rangle\no\\
&=&\prod_{r=\frac{1}{2}}\left(1+\frac{y}{r}\right)^2
\left[1-e^{-2r\tau}\left(\frac{r-y}{r+y}\right)^2\right] \no
\ee
where
\be
N_F = \sum_{r = \frac{1}{2}} r b^\dagger_r b_r, \quad
\tilde{N}_F = \sum_{r = \frac{1}{2}} r \tilde{b}^\dagger_r \tilde{b}_r
\ee
Thus, the total partition function may be written as
\be\label{superclose}
Z_S(y,\tau)&=& 2^2 Z_B(y,\tau) Z_F(y,\tau)\no\\
&=& 4T^2_p \sqrt{\frac{2\pi}{y+y^2\tau /2}}\prod_{n=1}
\left(1+\frac{2y}{n}\right)^{2}\left(1+\frac{y}{n}\right)^{-4}
\frac{1-e^{-2n\tau}\left(\frac{n-2y}{n+2y}\right)^2}
{\left[1-e^{-2n\tau}\left(\frac{n-y}{n+y}\right)^2\right]^2}
\ee
where we use a simple identity
\be\prod_{r={\frac{1}{2}}}f(r)
=\frac{\prod_{n=1}f(n/2)}{\prod_{n=1}f(n)}.
\ee
At a glance, as in the bosonic case, we find that, the partition
function evaluated in the boundary state formulation Eq.(\ref{superclose})
is the same as the partition function evaluated in the
BSFT Eq.(\ref{superopen}). Comparison between two results yields
\be
\tau = - \ln a, \quad c_S = 4\pi^{3/2}.
\ee

\section{Discussion}

In this paper we show that the open string one-loop partition
function in the BSFT can be calculated in the boundary state
formulation as the expectation value of the closed string
propagator between the boundary states in the presence of tachyon
condensation, off-shell. It is this equivalence that the
boundary state formulation is entirely based on. It is also this
equivalence that Polchinski \cite{polchinski} utilized to explore
the fundamental properties of the D-branes. However,
use of this equivalence has been limited to on-shell.
Recently the boundary state formulation has been proven to be
useful to discuss the tachyon condensation. So, it becomes
important to see that this equivalence holds for higher loop
partition function if we adopt the boundary state formulation
to explore the corrections to the effective tachyon action.
By an explicit calculation we show that it
holds for the open string one-loop partition function.
Extrapolating from the present work, we may propose that
all higher loop partition functions may be generated by the
following closed string field theory\cite{tlee2}, where the 
D-brane plays a role of source for the closed string field \cite{clsft}
\be \label{cls}
S = \frac{1}{2} \int \Phi {\cal K} \Phi + \frac{g^2}{3!} \int 
\Phi * \Phi * \Phi + \kappa \int \Phi J,
\ee
where
\be
{\cal K} &=& \hat{L}_0 + \hat{\bar L}_0, \no\\
\Phi &=& \Phi [x, {\bar x}], \no\\
J &=& e^{iS_{\rm bndy}[x, \bar{x}]}.\no
\ee
%We denote collectively the string coordinates in the longitudinal
%and transverse directions by $x_\parallel$ and $x_\perp$.
The source term in the closed string field action is written as
\be
\kappa \int \Phi J = \kappa \langle \Phi | B \rangle,
\ee
and the disk partition function appears as a normalization factor of the
boundary state when it is written in a form of coherent state
\be
|B \rangle &=& \int D[x, \bar{x}] e^{iS_{\rm bndy}[x, \bar{x}]}
|x, \bar{x} \rangle \\
&=& Z_{Disk} \exp \left(a^\dagger M \tilde{a}^\dagger
\right) |0 \rangle. \no
\ee
As we have seen that the open string one-loop partition function
corresponds to the closed string propagator. It is interesting to see
that all open string higher loop diagrams are generated as tree
level diagrams of this closed string field theory. In addition to
the open string loops, we find that the closed string field action
Eq.(\ref{cls}) generates closed string loops, i.e., the string
diagrams with handles. Figure 2. depicts a string diagram with one
handle.
\begin{figure}
\epsfxsize=6cm \centerline{\epsffile{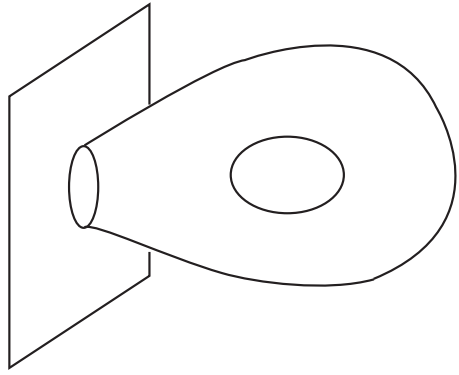}} \vspace{0cm}
\caption{String diagram with one handle}
\label{handle}
\end{figure}
It should be appropriate also to take into account the string diagrams
with handles when we calculate corrections to the disk
partition function. The closed string field theory we propose here
Eq.(\ref{cls}) generates the string diagrams with arbitrary number
of boundaries and handles, of which boundaries are attached on the
D-brane. Hence, it may serve as a consistent
framework to calculate corrections Eq.(\ref{total}) to the disk
partition function systematically. One of the good features of
the closed string field theory is that we only need to deal
with the weak coupling regime in order to discuss the tachyon
condensation in contrast to the open string field theory
\cite{tlee3}.

We conclude this paper with a remark on the infrared fixed point
limit. In the infrared fixed limit, $u \rightarrow \infty$, the
unstable Dp-brane turns into a lower dimensional D-brane. It has
been shown at the tree level by explicit calculations of the
partition function on a disk \cite{tlee1}. From the explicit
calculation given in this paper we also observe this phenomenon at
open string one-loop level. It follows from Eq.(\ref{close}) that
the partition function of the bosonic string theory reduces to \be
\lim_{u \rightarrow \infty} Z_B(u, \tau) = T^2_p
\frac{4\pi\sqrt{\pi}}{\sqrt{\tau}} \prod_{n=1} \frac{1}{1-
e^{-2n\tau}} \ee in the infrared fixed point limit. Since the
propagator of the zero mode with initial and final points fixed is
given as \be \langle x| e^{-\tau \hat{p}^2} | x \rangle =
\frac{1}{2 \sqrt{\pi \tau}}, \ee the infrared fixed point limit of
$Z_B$ can be understood as \be \lim_{u \rightarrow \infty}
Z_B(u,\tau) = (8\pi^2 T^2_p) \langle D| e^{-\tau(L_0 +
\tilde{L}_0)} |D \rangle \ee where $|D \rangle$ corresponds to the
boundary state with the Dirichlet boundary condition. Thus, the
one-loop partition function of the Dp-brane reduces to that of
D(p-1)-brane. Then it implies that $8\pi^2 T^2_p = T^2_{p-1}$,
i.e., the descent relation between D-brane tensions \be T_{p-1} =
2\pi \sqrt{\alpha^\prime} T_p. \ee The infrared fixed point limit
of the open string one-loop partition function of the superstring
Eq.(\ref{superclose}) can be discussed in a similar way. The 
present work can be also extended to the noncommutative string 
theories \cite{tlee1,noncomm,tlee4}.

\vskip 0.5cm

\noindent {\bf Acknowledgements}: This work is supported by a
grant from Natural Sciences and Engineering Research Council of Canada.
The work of TL was supported by KOSEF (995-0200-005-2) and was
done during the PIMS-APCTP Summer Workshop 2001 held at Simon-Fraser
University, Canada.

\end{document}